 \documentclass[a4paper]{article}
 \usepackage{graphicx}
 \begin{document}

 \title{T$_{c}$ map and superconductivity of simple metals at high pressure}
 \author{Wei Fan, Y. L. Li, J. L. Wang, L. J. Zou and Z. Zeng \\
 Key Laboratory of Materials Physics, \\
 Institute of Solid State Physics,
 Hefei Institutes of Physical Sciences, \\
 Chinese Academy of Sciences, 230031-Hefei,
 People's Republic of China}
 \date{\today}
 \maketitle
\begin{abstract}
We calculate T$_{c}$ map in region of weak electron-phonon coupling
based on simple phonon spectrum. By using linear-response method and
density functional theory, we calculate phonon spectra and Eliashberg
functions of simple metals under pressure. Based on the evolutions of
superconducting parameters of simple metals on the T$_{c}$ map with
increasing pressure, we find that there are two different responses to
pressure for simple metals: (1) enhancing electron-phonon interaction
$\lambda$ such as for La and Li, (2) increasing phonon frequency such
as for Pb, Pt. The $\lambda$ threshold effect is found, which origins
from the competition between electron-phonon interaction and
electron-electron Coulomb interaction and is the reason why T$_{c}$ of
most superconductors of simple metals are higher than 0.1K.
\end{abstract}

\begin{center}

\end{center}


\section{\label{intro}Introduction}

The BCS theory and its strong-coupling generalization (Eliashberg
theory) have successfully explained the superconductivity of
elementary metals and their alloys~\cite{Allen1,McMillan1}. In strong
electron-phonon coupling region $\lambda\gg\mu^{*}$, Eliashberg
equation has been widely studied, where $\lambda$ is the parameter of
electron-phonon interaction and $\mu^{*}$ is Coulomb pseudo-potential.
However in weak-coupling region $\lambda$ is comparable with
$\mu^{*}$. The competitions between electron-phonon interaction and
electron-electron interaction are so prominent that some new aspects
of Eliashberg equation can be found in such region. The Eliashberg
equation is very hard to be solved using analytic method. In fact no
analytic solution has been found except for some extreme limit. The
numerical methods can overcome the difficulties encountered by
analytic method and provide full and global information on Eliashberg
equation. The T$_{c}$ maps in three-dimensional parameter space
$\Omega-\lambda-\mu^{*}$ had been calculated~\cite{Fan1}, which can
predict the possible highest transition temperature for a certain
superconducting material and light the route to find superconductors
with higher T$_{c}$~\cite{Fan1}. It's very interesting to study the
evolution of superconducting parameters on the T$_{c}$ map with
external parameter such as pressure.

The issues on the superconductivity of simple elements become active
due to the developments of high pressure technique based on
diamond-anvil cell (DAC). The T$_{c}$ of metal lithium can reach about
20K at high pressure~\cite{Shimizu1,Struzhkin0}. The researches along
this direction have been extended to non-metal elements, such as
sulphur~\cite{Struzhkin1}, oxygen~\cite{Shimizu2} and
diamond~\cite{Ekimov1}, non-magnetic metals such as non-magnetic
iron~\cite{Shimizu3}, calcium~\cite{Yabuuchi1} and
yttrium~\cite{Hamlin1}. Some of them have very complex lattice
structures, rich structural phase diagrams and corresponding
structural phase transitions. Thus from structural complexity point of
view, these simple elements, such as lithium and boron, are not really
simple. The T$_{c}$ of these superconductors generally change with the
changes of crystal structures. At very high pressure some simple
metals even transform to insulator phase or semiconductor
phase~\cite{Ma1,Matsuoka1}. An introduction for the superconductivity
of simple elements can be found in a short review~\cite{Buzea1}.

Theoretically, the linear response method in the framework of density
functional theory has been widely used to calculate the
superconducting properties of simple elements~\cite{Savrasov1,Bose1}.
These calculations reproduce very well the transition temperatures of
some elements such as metal lithium~\cite{Kasinathan1,Christensen1},
yttrium~\cite{Yin1} and lanthanum~\cite{Pickett1,Tutuncu1} at high
pressure. In this paper we present the information on so-called
T$_{c}$ map~\cite{Fan1} in weak-coupling region. The numerical
calculations in weak-coupling region are more difficult than those in
strong-coupling region because the very low T$_{c}$ needs more
Matsubara frequencies to obtain more numerically reliable T$_{c}$. In
this work, we find the threshold effect of $\lambda$, which can
explain why noble metals are not superconductors even at very high
pressure. We use density functional theory and perturbed line-response
theory to calculate the Eliashberg function $\alpha^{2}F(\omega)$. We
mainly concentrate on the non-magnetic simple metals and try to find
how their T$_{c}$ run on the T$_{c}$ map with increasing pressure.
With increasing pressure, two different pressure effects are found:
(1) enhancing $\lambda$ such as for La and Li and (2) increasing
phonon frequency such as for Pb and Pt. The metals belong to the first
group generally undergo complex structural phase transitions with
increasing pressure.

\section{\label{sct}Strong-Coupling theory}

The energy-gap equation of Eliashberg-Nambu theory in the Matsubara's
imaginary-energy form is standard and well established~\cite{Allen1}.
Firstly, we use a simple analytic model of Eliashberg function written
in the form
 \begin{equation}\label{Eliashberg}
 \alpha^{2}F(\omega)=\left\{
 \begin{tabular}{cc} $\frac{c}{(\omega-\Omega_{P})^{2}+(\Omega_{2})^{2}}
  -\frac{c}{(\Omega_{3})^{2}+(\Omega_{2})^{2}}$, &
  $|\omega-\Omega_{P}|<\Omega_{3}$ \\ 0 & others, \end{tabular}\right.
 \end{equation}
\noindent where $\Omega_{P}$ is the energy  (or frequency) of phonon
mode, $\Omega_{2}$ the half-width of peak of phonon mode and
$\Omega_{3}=2\Omega_{2}$. The parameter of electron-phonon
interaction can be written as
$\lambda=\lambda(0)=2\int_{0}^{\infty}d\omega\alpha^{2}F(\omega)/\omega$.
The moments $\langle\omega^{n}\rangle$ of the distribution function
$(2/\lambda)\alpha^{2}F(\omega)/\omega$ are defined as
$\langle\omega^{n}\rangle=2/\lambda\int^{\infty}_{0}d\omega\alpha^{2}F(\omega)\omega^{n-1}$.
Based on these moments, others frequency parameters are defined as
$\langle\omega\rangle_{n}=(\langle\omega^{n}\rangle)^{1/n}$. Another
important parameter is $\langle\omega\rangle_{ln}=\exp[2/\lambda\int
(d\omega/\omega)\alpha^{2}F(\omega)\ln\omega]$.  The Coulomb
pseudo-potential is defined as $\mu_{0}=N(0)U$ and its renormalized
value is $\mu^{*}=\mu_{0}/(1+\mu_{0}\ln(E_{C}/\omega_{0}))$, where
$U$ is the Coulomb parameter, $N(0)$ the density of states at Fermi
energy, $E_{C}$ the characteristic energy for electrons such as the
Fermi energy or band width, and $\omega_{0}$ the characteristic
phonon energy such as the energy cutoff of phonon energy or Debye
energy. In the region of small $\lambda$, where T$_{c}$ is generally
smaller than 10K. To accurately define T$_{c}$, the number $N$ of
Matsubara energy must be larger enough to guarantee the range of
energy larger than the maximum of phonon energy. We choose $N$=1000,
the values of T$_{c}>$0.1K are numerically reliable if
$\Omega_{P}<$60 meV.

There are two important T$_{c}$ formulas based on strong coupling
theory, which are the McMillan formula\cite{McMillan1}
\begin{equation}\label{McMTc}
 T_{c}=\frac{\langle\omega\rangle_{1}}{1.20}
 \exp[-\frac{1.04(1+\lambda)}{\lambda-\mu^{*}(1+0.62\lambda)}]
\end{equation}
and the formula generalized by Allen~\cite{Allen1}
\begin{equation}\label{AllenTc}
 T_{c}=\frac{f_{1}f_{2}\langle\omega\rangle_{ln}}{1.20}
 \exp[-\frac{1.04(1+\lambda)}{\lambda-\mu^{*}(1+0.62\lambda)}],
\end{equation}
\noindent where $f_{1}$ and $f_{2}$ are the functions of $\lambda$,
$\mu^{*}$, $\langle\omega\rangle_{2}$ and $\langle\omega\rangle_{ln}$,
which can be found in Ref.\cite{Allen1}.

\begin{figure}
\begin{center}\includegraphics[width=0.80\textwidth]{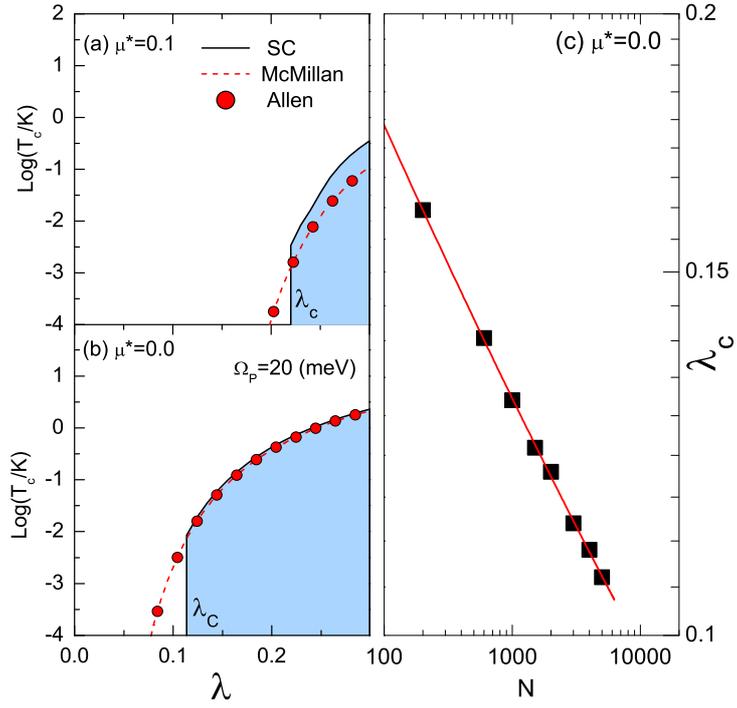}
\caption{\label{fig1}The comparison of strong-coupling theory with the
McMillan's formula and the Allen's T$_{c}$ formula. Two values of
Coulomb pseudo-potential (a) $\mu^{*}$=0.1 and (b) 0.0 are chosen but
the energy of phonon $\Omega_{P}$=20 meV keeps unchanged. The figure
shows the jump of T$_{c}$ when $\lambda$ is smaller a certain
threshold value $\lambda_{c}$ which is about 0.22 when $\mu^{*}$=0.1.
The (c) shows the N-dependent $\lambda_{c}$ plotted with
logarithm-scale axes.}
\end{center}\end{figure}

\begin{figure}
\begin{center}\includegraphics[width=0.80\textwidth]{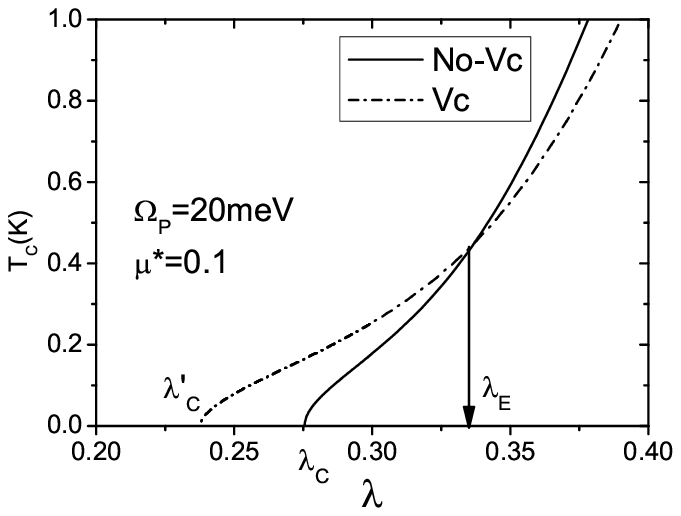}
\caption{\label{fig2} The effects of vertex corrections on
$\lambda_{c}$ using the same parameters as the Fig.\ref{fig1}(a). The
arrow shows the transition from negative vertex correction to positive
vertex correction. The band-width parameter E$_{B}$=1 eV is used in
calculations. }
\end{center}
\end{figure}

\section{\label{Threshold} Threshold effect of $\lambda$}

The McMillan's formula presents a threshold
$\lambda_{a}=\mu^{*}/(1-0.62\mu^{*})>\mu^{*}$. If
$\lambda<\lambda_{a}$, the superconductivity will be destroyed by
Coulomb interaction. A solution of Eliashberg equation had been found
long time ago by Wu and his co-workers~\cite{Wu1,Gong1,Wu2}. The
formula is expressed as
$T_{c}=\alpha_{00}\sqrt{\lambda\langle\omega^{2}\rangle}S(1/\lambda)$,
where $S(1/\lambda)=\sum_{j=0}^{\infty}b_{j}/\lambda^{j}$. The formula
is accurate in strong-coupling region. If number of Matsubara energies
used in calculation is large enough the formula is analytic when
$\infty>\lambda>\lambda_{a}$ even with $\lambda_{a}<$1.0. It had been
proved that if $N\gg$1, $\lambda_{a}\sim\mu^{*}$~\cite{Gong1}. When
$\lambda<\lambda_{a}$, the T$_{c}$ formula is not analytic and
physically, superconductivity does exist. If the number N of Matsubara
energies is large, the form of the T$_{c}$ formula will be very
complicated.

What is the threshold in general Eliashberg theory?. In this section
we present our numerical solution by using large number of Matsubara
energies up to N=5500. From Fig.\ref{fig1}(a), we find that, for
$\lambda<0.3$, the results of strong-coupling calculations are
considerably different from McMillan's and Allen's formulae. The
strong-coupling theory used in this work shows a disrupted change for
T$_{c}$.  When $\lambda<\lambda_{c}\sim 0.22-0.26$, there are no
superconductivity. The $\lambda_{c}$ is dependent on the number of
Matsubara energies. In our calculations, we increase the number from
200 to N=5500, the $\lambda_{c}$ decreases from 0.26 to 0.22 with
$\mu^{*}=0.1$. If we use larger $\mu^{*}$, $\lambda_{c}$ also takes
larger values. In Fig.\ref{fig1}(b), we decrease $\mu^{*}$ to zero,
the $\lambda_{c}$ decreases to about 0.10-0.11. The $\lambda_{c}$ in
many aspects are similar to $\lambda_{a}$ obtained from McMillan
formula, for instance, $\lambda_{c}$ slightly larger that $\mu^{*}$.
This means that $\lambda_{c}=\mu^{*}+\delta\lambda$ and
$\delta\lambda$$>$0. If $\mu^{*}$=0, $\delta\lambda=$0.1$\sim$0.11
shown in Fig.\ref{fig1}(b).

The accurate $\lambda_{c}$ is hard to define numerically because a
very large N is needed to obtain reliable results if T$_{c}<$0.0001K.
This brings difficulties to define T$_{c}$ if T$_{c}<$$10^{-4}$K using
numerical calculations of Eliashberg theory. At this time, McMillan
formula is a good choice as shown in Fig.\ref{fig1}(a,b). In the case
of $\mu^{*}$=0.0, the Fig.\ref{fig1}(c) show N-dependent $\lambda_{c}$
with logarithm-scale. we fit N-dependent $\lambda_{c}$ using formula
$\lambda_{c}=AN^{-P}+\lambda_{c}^{0}$. The fitting results are
$P$=-0.16 and $\lambda_{c}^{0}$=0.0299. This means that, if
$N\rightarrow\infty$, $\lambda_{c}\rightarrow\lambda_{c}^{0}$=0.0299.
Thus our results imply that the threshold of $\lambda$ exists, but not
so large as shown in Fig.\ref{fig1}. Additionally, there are many
physical reasons that make $\lambda_{c}$ not too small. In fact, only
finite $N$ is enough because of the existence of the cutoff for phonon
energy (Debye energy). Moreover some pair-breaking interactions such
as the interaction with magnetic order and magnetic fluctuation make
$\lambda_{c}$ larger.

We further study the influence of vertex correction or non-adiabatic
effects on $\lambda_{c}$. Generally, we expect that in strong-coupling
region with $\lambda>2$, the vertex correction has significant
influence on T$_{c}$. However in weak-coupling region from
Fig.\ref{fig2}, we find that the vertex correction can reduce
$\lambda_{c}$ to smaller value. This is closely related to the
positive vertex correction $T^{V}_{c}>T_{c}$ ($T^{V}_{c}$ is
transition temperature including vertex correction) that had been
found in previous work~\cite{Fan1}. The magnitude of vertex correction
is dependent on the effective band-width, in which the smaller
band-width is corresponding to the strong vertex correction. In
Fig.\ref{fig2}, the effective band-width is about 1.0 eV. The band
widthes of simple metals are generally larger than 1.0 eV so that the
vertex correction has smaller effects than that shown in
Fig.\ref{fig2}.

We have found that both larger N and vertex correction make
$\lambda_{c}$ decrease to the small value. As an approximation, we
define an approximate threshold $\lambda_{c}$=0.22-0.28 if
$\mu^{*}$=0.1. The region with $\lambda<\lambda_{c}$ and T$_{c}<$0.1K
is called weak-superconductivity region. The superconductivity in this
region is easily destroyed by pair-breaking interaction coming from
magnetism or inhomogeneity of superconducting materials. In fact, for
most of simple metals, T$_{c}$ are larger than 0.1K except for W, Be ,
Pt and Rh.

T$_{c}$ changing discontinuously with $\lambda$ will have significant
effects on the superconductivity. Simple metals have very low T$_{c}$
and their parameters of electron-phonon interaction $\lambda$  are
generally smaller than 0.5. The threshold effect can predict why some
noble metals such as Au, Ag and Cu are not superconductors even under
very-high pressure. The threshold effect determines why  T$_{c}$ of
most of simple metals are higher than 0.1K. We will extend the
Fig.\ref{fig1} to T$_{c}$ map on $\langle\omega\rangle_{2}-\lambda$
plane (Fig.\ref{fig4}) in the section \ref{TcMap}.

\begin{table}
\caption{\label{tab1} Table.(1)  The table provides the parameters of
lattice constant $a$,$b$,$c$ ($\AA$), $\lambda$ electron-phonon
interaction, $\langle\omega^{2}\rangle\lambda$ (meV)$^{2}$, transition
temperature T$_{c}$ (K) calculated in this work and the experimental
T$_{c}^{exp}(K)$. The numbers in parentheses behind T$_{c}$ and
T$_{c}^{exp}$ are pressure. For the purpose of comparison the
parameters $\lambda$ obtained in other works are also provided. The
second column is lattice parameters. }
 \begin{center}
 \begin{tabular}{llrrrrl}
 \hline\hline
  &  &$\lambda$&   & $\langle\omega^{2}\rangle\lambda$ & T$_{c}$ & T$_{c}^{exp}$ \\
 \hline
 Ni &$a$=3.51& 0.15 &                  &  101  &  -   &    -         \\
 Pd &$a$=3.97& 0.20 & 0.350$^{c}$      &  58.55  &  -   & 3.20$^{a}$ \\
 Pt &$a$=4.00& 0.32 &                  &  53.2  & 0.15 & 0.02$^{b}$ \\
 Cu &$a$=3.62& 0.17 & 0.140$^{c}$      &  50.12  &  -   &    -         \\
 Ag &$a$=4.15& 0.12 & 0.146$^{e}$      &  23.70  &  -   &    -         \\
 Au &$a$=4.13& 0.21 & 0.170$^{d}$      &  19.16  &  -   &    -         \\
 Cd & $a$=2.97 & 0.55 & 0.710$^{e}$      &  31.8  & 0.98 & 0.52         \\
    & $c/a$=1.88&&&&& \\
 Hg$^{h}$ &$a$($b$,$c$)=3.18& 1.85 &            &  21.1  & 4.87 & 4.15         \\
          &$\alpha(\beta,\gamma)=70.52^{\circ}$&&&&& \\
 Pb &$a$=4.77& 1.41 & 1.680$^{c}$      &  50.0  & 8.01 & 7.20         \\
 Li$^{f}$ &$a$=3.38& 0.33 &            &  202.2 & 22(41) & 20(50) \\
 Cs &$a$=6.13 & 0.12 &                  &  0.95  &  -   & 1.66(8)      \\
 La$^{g}$ &$a$=5.34& 0.986 &             &  63.24 & 6.84 & 6            \\
 Al &$a$=3.95& 0.37 & 0.440$^{c}$      &  352.2 & 1.20 & 1.18         \\
 Tl &$a$=3.45 & 1.58 &                  &  25.4  & 5.20 & 2.40         \\
    &$c/a$=1.59&&&&& \\
 \hline\hline
 \end{tabular}

 $^{a}$ Stritzker et al., irradiation with He$^{+}$ Ref.\cite{Stritzker1};
 $^{b}$ Schindler et al., powder samples, Ref.\cite{Schindler1};
 $^{c}$ Savrasov et al., Ref.\cite{Savrasov1};
 $^{d}$ Bauer et al., Ref.\cite{Bauer1};
 $^{e}$ Bose, Ref.\cite{Bose1} with $\lambda$  of 3d and 4d transition metal elements;
 $^{f}$ fcc structure. At ambient pressure Li is metal with bcc structure
 without superconductivity.
 $^{g}$ fcc structure.
 $^{h}$ Rhombohedral structure.
 \end{center}
 \end{table}

\begin{figure}\begin{center}\includegraphics[width=0.80\textwidth]{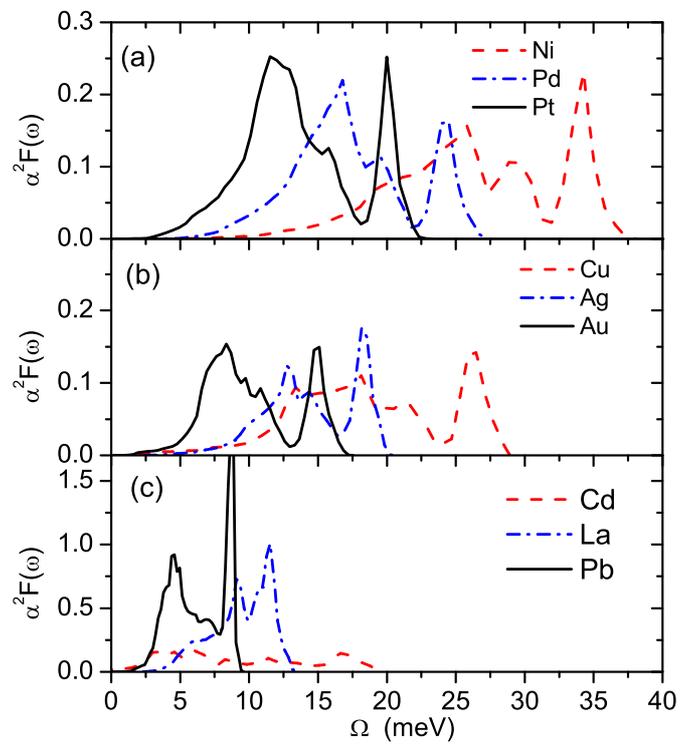}
\caption{\label{fig3} The Eliashberg-McMillan functions
$\alpha^{2}F(\omega)$ for simple elements of Ni, Pd, Pt, Cu, Ag, Au,
Cd, La and Pb. }
\end{center}\end{figure}

\section{\label{dft_phonon} The phonon spectrum and electronic-phonon interaction from
the calculations of linear response method}

With development of the method of electronic-structure calculation in
solid materials, the Eliashberg function $\alpha^{2}F(\omega)$ and the
parameters of electron-phonon interaction can be calculated using
density functional theory combined with linear-response
theory~\cite{Savrasov1}. The q-dependent parameter of electron-phonon
interaction is defined as
\begin{equation}\label{lamq}
 \lambda_{q\nu}=\frac{1}{\pi N_{s}(0)}\frac{\gamma_{q\nu}}{\omega^{2}_{q\nu}},
\end{equation}
\noindent where $\gamma_{q\nu}$ is phonon linewidth, $\omega_{q\nu}$
the dispersion relation of phonon, $N_{s}(0)$ the electronic density
(pre spin) at Fermi surface. The usual $\lambda$ is the weighted
average of $\lambda_{q\nu}$ for different $q$ points and the summation
of all phonon bands or phonon modes. In this work, the
$\alpha^{2}F(\omega)$, $\gamma_{q\nu}$ and $\lambda_{q\nu}$ are
calculated using the linear response theory with the frame of the
density functional theory based on the plane-wave pseudo-potential
method~\cite{pwscf1}.  We just simply use the q-independent Eliashberg
function $\alpha^{2}F(\omega)$ in standard Eliashberg theory. A full
treatment electron-phonon interaction needs q-dependent Eliashberg
function $\alpha^{2}F(k,\omega)$ introduced in
reference~\cite{Allen2}.

The ultra-soft pseudo-potential and the GGA
approximation~\cite{Perdew1} are used in total-energy calculation of
electronic structure~\cite{pwscf1}. In the calculations, two
Monkhorst-Pack grids 16$\times$16$\times$16 and 24$\times$24$\times$24
are used for fcc and bcc structure, two grids of 16$\times$16$\times$8
and 24$\times$24$\times$12 for hcp structure. The Monkhorst-Pack grids
for the phonon calculations are 4$\times$4$\times$4 and
6$\times$6$\times$6 for fcc and bcc, 4$\times$4$\times$2 and
6$\times$6$\times$3 for hcp structures respectively. The energy
cutoffs are different for different elements within range from 400 eV
to 550 eV.

Using the Eliashberg function $\alpha^{2}F(\omega)$ obtained from
linear response theory and the standard strong-coupling theory, we can
more reliably calculate T$_{c}$ and other properties of
superconductors. We have calculated the Eliashberg functions
$\alpha^{2}F(\omega)$ of simple metals listing in Table.(\ref{tab1}).
We plot $\alpha^{2}F(\omega)$ functions of Ni, Pd and Pt in
Fig.\ref{fig3}(a), Cu, Ag and Au in Fig.\ref{fig3}(b), Cd, La and Pb
in Fig.\ref{fig3}(c). From Table.(\ref{tab1}), metallic Ni has
relatively large hopfield parameter $\langle\omega^{2}\rangle\lambda$,
so it is a potential superconductor under extreme condition such as at
high pressure similar to non-magnetic iron~\cite{Shimizu3}. Our
calculations have shown that the magnetic moment of Ni atom quenches
to zero when pressure is higher than 300 GPa. The element Pd has
smaller $\lambda\sim 0.20$ and its superconductivity is only found in
the samples after the irradiation with He$^{+}$~\cite{Stritzker1}. The
element Pt has relative larger $\lambda$ value than the threshold
$\lambda_{c}$, so it should be a superconductor. The superconductivity
of element Pt, however, is only found in powder
samples~\cite{Schindler1}. From Table.(\ref{tab1}), the $\lambda$
values of noble metals Cu, Ag and Au are all smaller than the
threshold $\lambda_{c}\sim 0.22-0.28$, so the threshold effect for
small $\lambda$ in strong-coupling theory is indicative that noble
metals Cu, Ag and Au have no superconductivity. It's correct even at
high pressure (see next section). In Table.(\ref{tab1}), the elements
of heavy metals such as Pb, Hg, Cd and La have relative large
$\lambda$. For metallic Pb, we use the norm-conservation
pseudo-potential which gives better phonon spectrum and T$_{c}$. The
Hg and Cd have smaller $\langle\omega^{2}\rangle\lambda$ values so
these heavy metals have small space to increase T$_{c}$ further at
extreme condition such as high-pressure. The T$_{c}$ in
Table.(\ref{tab1}) for Cd, Hg, Pb and La match the experimental values
very well. These heavy metals have soft modes with imaginary energies
at normal condition. This agrees with general arguments that the
softening of vibration mode is intrinsic important to
superconductivity. It's reasonable that we use the Coulomb
pseudo-potential $\mu^{*}$$\sim$0.1 in strong coupling calculation for
all simple metals under studied. This is because the obtained T$_{c}$
are very close to experiments shown in Table.\ref{tab1}. The
underlying physical reason is probably that large band-width of these
simple metals make the re-normalized $\mu^{*}$ very close to 0.1.

\begin{figure}
\begin{center}\includegraphics[width=0.80\textwidth]{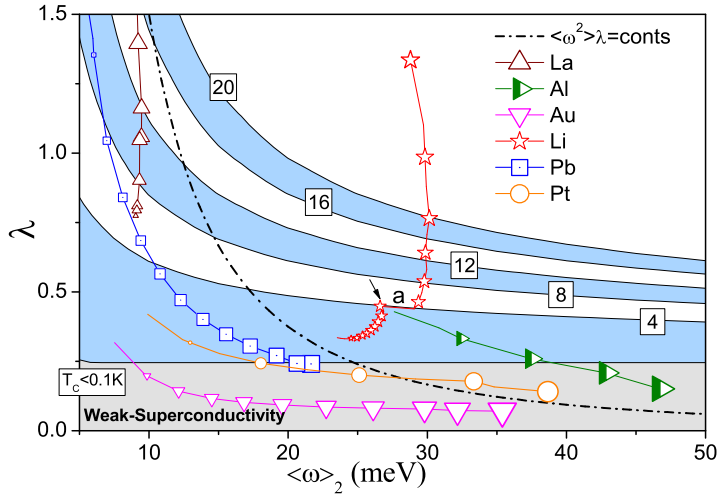}
\caption{\label{fig4} The evolutions of T$_{c}$ with superconducting
parameters $\lambda$ and $\langle\omega\rangle_{2}$ on T$_{c}$ map for
different metals with increasing pressure. The sizes of scatters in
proportion to $P^{1/5}$ measure the magnitude of pressure $P$. The
arrow $a$ points the break of curve for lithium at 13 GPa.  The
dash-dot line plot the curves $\langle\omega^{2}\rangle\lambda$=cont.
}
 \end{center}\end{figure}

\begin{figure}
\begin{center}\includegraphics[width=0.80\textwidth]{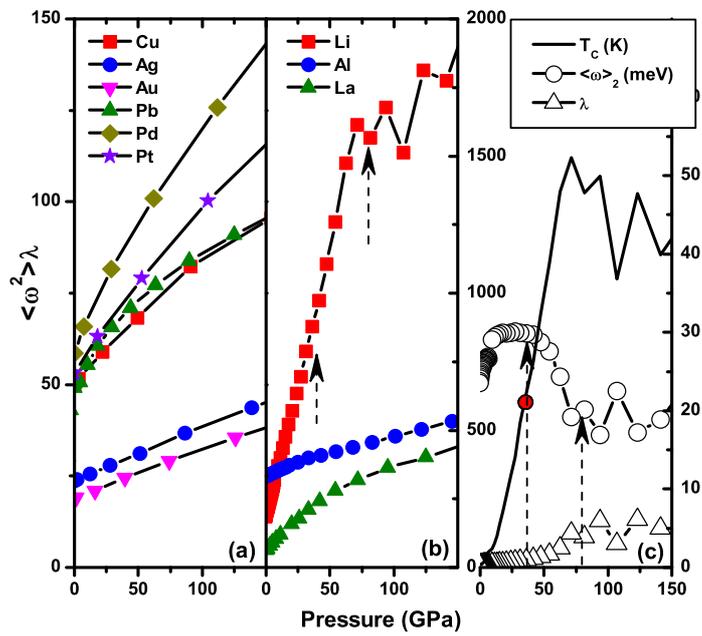}
\caption{\label{fig5} (a,b) The parameters
$\langle\omega^{2}\rangle\lambda$ with unit (meV)$^{2}$ change with
increasing pressure for different simple metals. (c) T$_{c}$,
$\lambda$ and $\langle\omega\rangle_{2}$ of metal Li change with
increasing pressure. The two arrows in (b) and (c) respectively are
corresponding to the structural phase-transitions of metal Li at
pressures 40 GPa and 80 GPa. }
\end{center}\end{figure}

\section{\label{TcMap} The evolution of superconductivity with increasing pressure on T$_{c}$ map}

As main purpose of this context, we study the evolutions of
superconducting state on T$_{c}$ map~\cite{Fan1} when pressure
increases. The static pressure is simply controlled by changing the
cell volume. We plot the evolutions of superconducting parameters of
Al, Au, Pt, Li, La and Pb on the map with increasing pressure shown in
Fig.\ref{fig4}. The $\langle\omega\rangle_{2}$ is used to measure the
phonon energy instead of $\Omega_{P}$ used in previous
papers~\cite{Fan1}. The $\lambda$ for Au decreases with pressure and
$\langle\omega\rangle_{2}$ increases with pressure, which is very
similar to others noble metal Cu and Ag. The $\lambda$ keeps smaller
than $\lambda_{c}$ is indicative that Au, Ag and Cu can not become
superconductors within pressure ranges up to 300GPa. At ambient
pressure, Pt is marginal superconductor, but its superconducting
parameters enter non-superconductivity region when increasing
pressure. The metal Al is real superconductor at ambient pressure and
becomes non-superconductivity with increasing pressure. The metal Pb
is a strong-coupling superconductor and T$_{c}$ decreases with
increasing pressure.

The positive $dT_{c}/dP$ for La and Li with increasing the pressure
are different from the negative $dT_{c}/dP$ for Al, and Pb. For La and
Li, the parameters of electron-phonon interaction $\lambda$ increase
with pressure and the $\langle\omega\rangle_{2}$ changes relative
small (even decrease). A structural instability happens at 13 GPa for
Li. We recalculate the curve using norm-conserving pseudo-potential.
The structural instability still remains however it becomes very weak.
For Li, the ultra-soft pseudo-potential gives T$_{c}$=22K at pressure
36 GPa and for normal-conserving pseudo-potential T$_{c}$=20K at 15
GPa. Thus the results of ultra-soft pseudo-potential are more close to
experimental T$_{c}$=20K at pressure 50 PGa in Ref.~\cite{Shimizu1}
and T$_{c}$=16K at 40 GPa in Ref~\cite{Struzhkin0}. Similar to Li,
both $\lambda$ and T$_{c}$ of La increase with pressure and
$\langle\omega\rangle_{2}$ has only small change. The more sensitive
response of $\lambda$ than $\langle\omega\rangle_{2}$ for applying
pressure means that Li and La try to modify their lattice structures
under high pressure. However for Pb, Pt, Au and Ag under high pressure
the lattice structures are stable and the frequencies of phonons are
more efficiently increasing with pressure.

An important relation $\langle\omega^{2}\rangle\lambda$=constant has
been plotted on the T$_{c}$ map. The relation had been used to explain
the spatial anti-correlation between energy gap and phonon energy
found in the samples of HTSC superconductor Bi2212~\cite{Fan1} and the
anomalous isotope effects of Rb$_{3}$C$_{60}$~\cite{Fan2}. A very
useful formula of isotope effects had been derived~\cite{Fan2} based
on the relation. The pressure-dependent
$\langle\omega^{2}\rangle\lambda$ had been studied in detail using the
formula of strong-coupling theory~\cite{Chen1}. The
Fig.\ref{fig5}(a,b) show that $\langle\omega^{2}\rangle\lambda$ does
not keep as constant with increasing pressure.  However our
calculations show that the generalized relation
$M\langle\omega^{2}\rangle\lambda$=constant is correct for isotope
substitution for simple metals.

The curve $\langle\omega^{2}\rangle\lambda$ of Pb is very close to the
curves of Cu with non-superconductivity, Pt and Pd with weak
superconductivity shown in Fig.\ref{fig5}. The superconductivity of Pb
is because of the large $\lambda$ or strong-electron-phonon
interaction. The atomic mass of Pt is very close to Pb but with very
weak superconductivity. This is because the parameter $\lambda$ of Pt
is far less than that of Pb shown in Table.(\ref{tab1}), although the
phonon energy of Pt is almost twice higher than that of Pb shown in
Fig.\ref{fig3}. The parameter $\langle\omega^{2}\rangle\lambda$
measures the possible maximum of T$_{c}$, however the individual
values of $\langle\omega^{2}\rangle$ and $\lambda$ need to be
considered to determine the real T$_{c}$ of a superconducting
material. The parameter $\langle\omega^{2}\rangle\lambda$ increases
with pressure for most of superconductors in Fig.\ref{fig5}. However,
this does not mean that T$_{c}$ always increases with pressure. The
depressed T$_{c}$ with increasing $\langle\omega^{2}\rangle\lambda$
such as for Pb ascribes to the decrease of $\lambda$. For metal Al the
same situation is correct and its T$_{c}$ decreases with increasing
pressure and close to zero at 5 GPa~\cite{Gubser1}, which is
consistent with our calculations and others~\cite{Profeta1}.

Experimentally, metal Li shows complex structural phase-transitions in
diamond anvil when pressure increases up to 100 GPa.  There are at
least two phase transitions at around 40 GPa and around 70-80
GPa~\cite{Matsuoka1}. Around 20-40 GPa, fcc structure is stable.
T$_{c}$ reaches to 16K near 40GPa. From 40 GPa to 70-80 GPa the c/16
phase is stable and T$_{c}$ keeps about 16K~\cite{Struzhkin0} (even
reaches about 20K~\cite{Shimizu1}). When pressure is higher than 80
GPa, a structural phase happens and T$_{c}$ decreases. Compared with
our results shown in Fig.5(b,c), the phonon energy
$\langle\omega\rangle_{2}$ significantly deceases at pressure around
40 GPa, and with pressure increasing further, $\lambda$ reaches to the
maximum at pressure around 80 GPa. Thus our results mean that, between
40 GPa to 80 GPa, the crystal structure is in an intermediate phase
between two stable phases. At pressure 40 GPa, the fcc structure
becomes unstable, the intermediate phase emerges from fcc phase by
phonon softening. At pressure 70-80 GPa, the electron-phonon
interaction reaches to the maximum. The intermediate phase becomes
unstable and a new phase form by a phase transition. The big jump of
resistivity of normal state (25K) at about 80 GPa just shows a real
phase-transition~\cite{Matsuoka1}. The T$_{c}$ monotonously increases
with pressure from 40 to 75 GPa in our calculations, which is
different from keeping around 16K in experiment~\cite{Struzhkin0}. The
discrepancy is probably because the stable structure is the c/16 phase
and not the fcc structure used in our calculation.  When pressure
increases to higher than 80 GPa, T$_{c}$ decreases with pressure which
is consistent with experiments. Although the fcc structure is kept
unchanged from low to high pressure, our results still hint possible
structural transitions of metal Li at corresponding experimental
pressures. This means that the interactions of atomic orbits between
nearest-nearby atoms plays more important role than the formation of
super-structure of crystal with increasing pressure.

The pressure-dependent T$_{c}$ of metal Li is very similar to
superconducting niobium nitride (NbN). The observed high-pressure
effect on T$_{c}$ is explained as the pressure-induced interplay of
electronic stiffness and phonon frequency~\cite{Chen2}. In this paper,
the high-pressure effect on Li is explained as the changes of role
between electron-phonon interaction $\lambda$ and phonon energy
$\langle\omega\rangle_{2}$ when pressure increase as illustrated in
Fig.\ref{fig5}(c). In low-pressure region, increasing T$_{c}$ with
pressure is induced by the increasing $\lambda$. In high-pressure
($P>$80Gpa) region, decreasing T$_{c}$ with pressure is induced by
softening of phonon modes.

\begin{figure}
\begin{center}\includegraphics[width=0.80\textwidth]{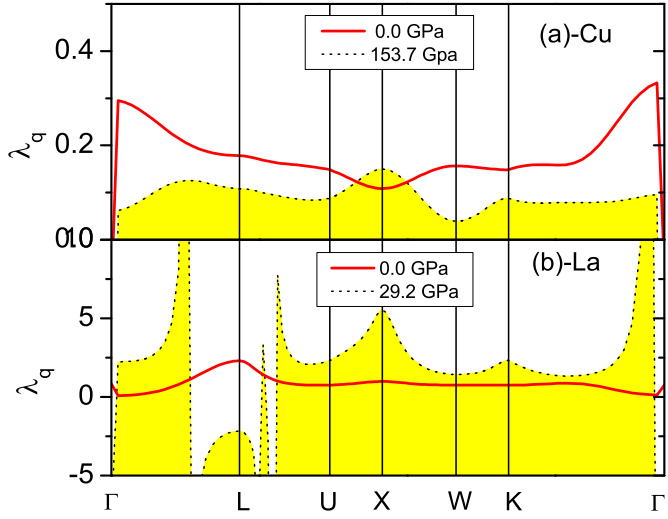}
\caption{\label{fig6} The q-dependent parameters $\lambda_{q}$ of
electron-phonon interaction at low pressure (solid line) and high
pressure (dot-line) 153.7 GPa (Cu) and 29.2 GPa (La). }
\end{center}\end{figure}

\begin{figure}
\begin{center}\includegraphics[width=0.80\textwidth,angle=0]{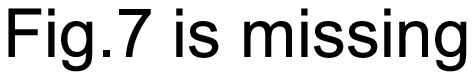}
\caption{\label{fig7} The Fermi surfaces of metal Cu at high pressure
153.7 GPa (a), and La (fcc) at low pressure 0.0 GPa (b) and high
pressure 29.2 GPa (b).}
\end{center}\end{figure}

\section{\label{pressure} The electron-phonon interaction of element metals under pressure}

The superconductors collected on T$_{c}$ map in Fig.\ref{fig4} can be
divided into two groups: (1) metals La and Li; (2) metals Pb, Al and
others. The effects of high pressure are more efficiently increasing
$\lambda$ for group (1) but increasing $\Omega_{P}$ for group (2). As
discussed in Ref.\cite{Buzea1} based on experimental T$_{c}$-pressure
relations, most of simple metals are belong to above two groups. We
calculate q-dependent $\lambda_{q}$ of La, Cu and other metals at
different pressures. As a comparison, we present the results of
non-superconductor Cu and superconductor La. The Cu is prefect
conductor and the metal La with fcc structure is well-known
superconductor at ambient pressure with T$_{c}$=6 K~\cite{Maple1}. At
high pressure 20 GPa, T$_{c}$ of La increases to 12.8K~\cite{Maple1}.
The electron-band, phonon-band and Eliashberg function
$\alpha^{2}F(\omega)$ of La had been studied in very
details~\cite{Pickett1,Tutuncu1}.

The q-dependent parameters $\lambda_{q}=\sum_{\nu}\lambda_{q\nu}$ of
electron-phonon interaction generally have very large values near or
at some high symmetry q-points shown in Fig.\ref{fig6}.  For metal Cu,
the $\lambda_{q\nu}$ decreases with increasing pressure but the
$\omega_{q\nu}$ increases with pressure. The one exception is that
near K point, $\lambda_{q}$ of Cu increases with pressure. The
increase of $\lambda_{q}$ at high symmetry point of Brillouin zone,
such as K point of fcc crystal, means that the C$_{2v}$ symmetry of
the crystal will be possible broken if pressure increases further. The
values of $\lambda_{q}$ for metal La have maximum near high-symmetry
points. Exactly at $\Gamma$ point, $\lambda_{q}$ is hard to define due
to the large numerical errors. However, main features at others $q$
points, especially the larger values of $\lambda_{q}$ near $\Gamma$
point along $\Gamma-K$ line are very robust in Fig.\ref{fig6}(b).

The phonon anomaly of fcc La at $L$ point shown in Fig.\ref{fig6}
were found in experiment of temperature-dependent phonon
dispersion~\cite{Stassis1} and theoretical
calculations~\cite{Tutuncu1}. The soft modes especially for
transverse modes (shearing instability) lead to the instability of
lattice. The distances (or interactions) between $\langle 111
\rangle$ planes are larger (weaker) than those along other
directions. The instability of lattice structure begins with
transverse gliding between $\langle 111 \rangle$ planes due to
smaller activation energy.

We plot the Fermi surfaces of La at low pressure and at high pressure
shown in Fig.\ref{fig7}. At low pressure there is only one sheet of
Fermi surface with rather complex topology. The electrons occupy
around the boundaries except that near the center region of eight
hexagons and the 24 vertexes. When the pressure increases to 29.2 GPa,
the topology of Fermi surface has significant changes. A group of very
small Fermi-surface packets appear near the middle of $\Gamma-K$ lines
but they aren't displayed in Fig.\ref{fig7} due to very small sizes.
The occupations of electrons near regions of eight hexagons have large
reductions, especially near 36 edges. This means that the electronic
states at the edges of Brillouin zone are electron-like states at low
pressure and hole-like states at high pressure. The large
reconstruction of Fermi surface for La will lead to the structural
transition between different phases. For metal Cu, the increasing
pressure only generates very small changes of lattice constant. So its
electron-like Fermi surface has very small change at high pressure
(153.7 GPa) shown in Fig.\ref{fig7} with the well-known shape of Fermi
surface at ambient pressure.

The high-symmetry points of Brillouin zone characterize the long-range
orders of crystal structure determined by space-group symmetries. The
electrons with wave vectors near high symmetrical points, line and
boundaries of Brillouion zone have potentially strong interactions
with lattice vibrations. The main contributions to $\lambda$ come from
the parts of Fermi surface contacted with boundaries of Brillouion
zone. This is why noble metals have very small $\lambda$ and metal La,
Pb and Hg have very larger $\lambda$. Thus, the structural phase
transition will lead to the sharp increases of $\lambda$ with
increasing pressure for metal La.

\section{\label{conclusion}conclusion}

In summary, we have calculated T$_{c}$ map in the region of weak
electron-phonon coupling. The threshold effect of $\lambda_{c}$ is
found in small $\lambda$ region. It origins from the competition
between electron-phonon interaction and electron-electron Coulomb
interaction when $\lambda$ is comparable with $\mu^{*}$. The threshold
effect of $\lambda$ can explain why noble metals are not
superconductors even at high pressure and why T$_{c}$ of most of
simple metals are higher than 0.1K. The phonon spectra and Eliashberg
functions have been calculated by using linear response method in the
framework of density functional theory. The obtained Eliashberg
functions $\alpha^{2}F(\omega)$ are then used in the calculations of
strong-coupling theory. Our calculations illustrate how the
superconducting parameters of simple metals distribute on the T$_{c}$
map and how they run on the map with increasing pressure. Our results
explain the pressure-dependent T$_{c}$ of metal Li at high pressure (
$P>$80GPa ). Thus the T$_{c}$ maps studied in this paper and in
previous papers~\cite{Fan1} are very useful tools for studying the
superconductivity of simple metals and hopefully extend to other types
of superconductors.

\section{\label{ack}Acknowledgements}

The Fermi surfaces are plotted using XCrySden software. Zeng was
supported by the special Funds for Major State Basic Research
Project of China (973) under Grant 2007CB925004 and Zou by the
National Science Foundation of China under Grant 10874186.

\end{document}